\documentstyle[pra,preprint,aps]{revtex}
\tightenlines
\begin{document}
\draft

\title{\bf GENERAL POLARIZATION MATRIX \\ OF
ELECTROMAGNETIC RADIATION}

\author{Muhammet Ali Can and Alexander S. Shumovsky}

\address{Department of Physics, Bilkent University, Bilkent, Ankara,
06533, Turkey}

\maketitle

\begin{abstract}
A general form of the polarization matrix valid for any type of
electromagnetic radiation (plane waves, multipole radiation etc.)
is defined in terms of a certain bilinear form in the
field-strength tensor. The quantum counterpart is determined as an
operator matrix with normal-ordered elements with respect to the
creation and annihilation operators. The zero-point oscillations
(ZPO) of polarization are defined via difference between the
anti-normal and normal ordered operator polarization matrices. It
is shown that ZPO of the multipole field are stronger than those
described by the model of plane waves and are concentrated in a
certain neighborhood of a local source.
\end{abstract}
\pacs{PACS No:42.25.Ja, 42.50.Lc}
 \narrowtext


\newpage

\section{Introduction}

It is well known that the polarization measurements play very
important role in optics and spectroscopy for a long time (e.g.,
see \cite{1}). At present, the polarization entanglement of photon
twins is widely discussed in connection with the fundamental
problems of quantum theory such as the nonlocality and existence
of hidden parameters. In spite of the fact that there is still a
certain question in regard to fundamental issues of the entangled
states of quantum systems, quantum entanglement has been
recognized as an important tool in quantum information processing
and quantum computation (e.g., see \cite{2} and references
therein).

The description usually given of the polarization is a classical
one, defining the polarization as a measure of transversal
anisotrophy of the plane electromagnetic waves \cite{1,3}. This
definition is based on the fact that the classical radiation field
measured in a small volume at far distance from a local source can
be approximated by the plane waves. In this case, the field
strengths $\vec E$ and $\vec B$ have only two symmetric spatial
components $E_x=B_y$ and $E_y=-B_x$ transversal with respect to
the direction of propagation chosen to be the $z$-axis. Since
these complex components may have different magnitude and phases,
the quantitative description of polarization is provided by the
so-called polarization matrix with the elements \cite{3}
\begin{eqnarray}
{\cal P}_{\sigma \sigma'}=E^*_{\sigma}E_{\sigma'}, \quad \quad
\sigma , \sigma' =x,y. \label{1}
\end{eqnarray}
Here $E_{\sigma}$ denotes the component of the positive-frequency
part of the classical electric field strength and $\vec B$ is the
magnetic induction. The quantum counterpart of (1) is represented
by the operator polarization matrix which is obtained from (1) by
a formal substitution of corresponding operators instead of the
complex field strengths \cite{4}.

The plane-wave approximation used in the definition of (1) and its
quantum counterpart is valid only at far distances from a local
source such as an atom or molecule. In the intermediate and near
zones, the field strengths can oscillate in any direction in spite
of the fact that ${\vec E} \cdot {\vec B}=0$ and $|E|^2-|B|^2=0$
everywhere. The atomic transitions emit the multipole radiation
represented by the spherical waves either classical or quantum
\cite{5}. In contrast to the plane waves, the field strengths of
the multipole field can have all three spatial components. For
example, the electric-type multipole radiation always has all
three components of the electric field strength, while the
magnetic induction obeys the condition of transversality ${\vec
B}({\vec r}) \cdot {\vec r}=0$. In contrast, the magnetic
multipole radiation has the three components of magnetic induction
at completely transversal electric field \cite{6}. Therefore, the
polarization of either multipole field is the three-dimensional
rather than two-dimensional property of the radiation. This means
that, instead of the $(2 \times 2)$ polarization matrix (1), we
have to consider a more general $(3 \times 3)$ polarization matrix
\cite{7}. In the plane wave approximation, this general
polarization matrix should be reduced to (1).

The three-dimensional structure of the polarization can be
illustrated as follows. The quantum theory interprets the
polarization as a given spin state of photons \cite{5}. The spin
of a photon is known to be $1$. Thus, it has just the three
independent spin states. In the case of plane waves of photons,
the third state with zero projection of spin on the axis of
quantization is forbidden and the polarization picture is reduced
to the transversal anisotrophy of the field. On the contrary, all
the three spin states are allowed in the case of multipole
radiation \cite{5}. Hence, in general case, we have to consider
the spatial anisotrophy of the field rather than the transversal
one. It is also clear that both the electric field strength and
magnetic induction can have nontrivial spatial properties and
therefore both of them should be taken into account \cite{8}.

The main scope of this note is to discuss a new general form of
the polarization matrix which can be defined directly in terms of
the field-strength tensor. We show, that the general polarization
matrix (GPM) fits the polarization of plane and multipole waves
adequately. We also construct the quantum counterpart of GPM and
discuss the zero-point oscillations of polarization.

\section{General polarization matrix}

It is well known that the general description of electromagnetic
field is based on the field-strength tensor  defined as follows
\cite{6}
\begin{eqnarray}
F({\vec r})= \left( \begin{array}{cccc} 0 & E_x & E_y & E_z \\
-E_x & 0 & -B_z & B_y \\ -E_y & B_z & 0 & -B_x \\ -E_z & -B_y &
B_x & 0
\end{array} \right) \label{2}
\end{eqnarray}
For example, the set of Maxwell's equations can easily be
expressed in terms of (2). Here the field strengths are determined
at the observation point $\vec r$.

To construct GPM, we assume that (2) consists of the
positive-frequency parts of the field strengths. In fact, $F$ is
additive with respect to the contributions coming from the
positive- and negative frequency parts. Consider the simplest
bilinear form
\begin{eqnarray}
R({\vec r})=F^+({\vec r})F({\vec r}), \label{3}
\end{eqnarray}
that can be constructed through the use of (2). It is clear that
this form differs from the energy-momentum tensor of the field by
a scalar. In some sense, (3) is similar to Ricci tensor considered
in general relativity \cite{9}. By construction, (3) is the $(4
\times 4)$ block matrix of the form
\begin{eqnarray}
R({\vec r})= \left( \begin{array}{cc} W_E & {\vec S} \\ {\vec S}^+
& P
\end{array} \right)  \nonumber
\end{eqnarray}
Here $W_E={\vec E}^* \cdot {\vec E}$ is the scalar, defining the
electric-field contribution into the energy density and ${\vec
S}={\vec E}^* \times {\vec B}$ is proportional to the Poynting
vector. We choose to interpret $P$ in (3) as GPM.

It is clear that $P({\vec r})$ is the Hermitian $(3 \times 3)$
matrix  additive with respect to contributions coming from the
electric field and magnetic induction:
\begin{eqnarray}
P({\vec r})=P_E({\vec r})+P_B({\vec r}), \label{4}
\end{eqnarray}
where
\begin{eqnarray}
P_E({\vec r})= \left( \begin{array}{ccc} E^*_xE_x & E^*_xE_y &
E^*_xE_z \\ E^*_yE_x & E^*_yE_y & E_y^*E_z \\ E^*_zE_x & E^*_zE_y
& E^*_zE_z
\end{array} \right) \label{5}
\end{eqnarray}
and
\begin{eqnarray}
P_B({\vec r})= \left( \begin{array}{ccc} B^*_yB_y+B^*_zB_z &
-B^*_yB_x & -B^*_zB_x \\ -B^*_xB_y & B^*_xB_x+B^*_zB_z & -B^*_zB_y
\\ -B^*_xB_z & -B^*_yB_z & B^*_xB_x+B^*_yB_y \end{array} \right)
\label{6}
\end{eqnarray}
By construction, the two terms in (4) describe the spatial
anisotrophy of the electric and magnetic fields respectively. The
diagonal terms give contribution of corresponding field strengths
into the intensity in given directions. The off-diagonal terms
provide the "phase information" about the phase difference between
the spatial components. For example, the off-diagonal terms in (5)
are specified by the two out of three phase differences
$\Delta_{\mu \mu'} = \arg E_{\mu'}- \arg E_{\mu}$ such that
\begin{eqnarray}
\Delta_{xy} + \Delta_{yz} + \Delta_{zx} =0.  \nonumber
\end{eqnarray}
In the case of plane waves described by the polarization matrix
(1), there is only one polarization phase difference \cite{3,6}.
The matrix (5) has been examined in \cite{7} in the case of
electric-type multipole radiation. In turn, (6) is similar to the
object has been proposed in \cite{8} for the magnetic-type
radiation.

Let us show that GPM (4) is reduced to (1) in the plane wave
representation. Assume that the waves propagate along the
$z$-direction. Then (5) takes the form
\begin{eqnarray}
P_E= \left( \begin{array}{cc} {\cal P} & 0 \\ 0 & 0
\end{array} \right) \nonumber
\end{eqnarray}
where ${\cal P}$ is the conventional polarization matrix (1). In
turn, (6) takes the form
\begin{eqnarray}
P_B= \left( \begin{array}{cc} {\cal P} & 0 \\ 0 & W_E \end{array}
\right) \nonumber
\end{eqnarray}
where $W_E={\vec E}^* \cdot {\vec E}$. Thus, in the case of plane
waves, GPM (4) fits the conventional theory of polarization of
plane waves adequately.

Consider now the case of electric-type multipole radiation when
$\forall {\vec r} \quad B_z({\vec r})=0$. Then, the matrix (6)
takes the form
\begin{eqnarray}
P_B({\vec r})= \left( \begin{array}{cc} {\cal P}^{(B)} & 0 \\ 0 &
W_B
\end{array} \right) \nonumber
\end{eqnarray}
where
\begin{eqnarray}
{\cal P}^{(B)}_{\sigma \sigma'}({\vec r}) \equiv
B^*_{\sigma}B_{\sigma'} , \quad \quad \sigma , \sigma' =x,y,
\nonumber
\end{eqnarray}
and $W_B={\vec B}^* \cdot {\vec B}$ is the magnetic contribution
into the energy density. Thus, in the case of electric-type
multipole radiation, the matrix (5) describes the spatial
anisotrophy of the electric field strength at the point $\vec r$,
while the magnetic contribution corresponds to the transversal
anisotrophy. Due to the orthogonality condition ${\vec E}({\vec
r}) \cdot {\vec B}({\vec r})=0$, there are only two independent
polarization phase differences $\Delta_{\alpha \beta}$ in this
case. The polarization of magnetic-type multipole radiation can be
described in terms of (4) in similar way.

It is well known that the contribution of radial component of the
multipole radiation decreases rapidly with the distance from a
local source \cite{6}. Therefore, at far distances,  the terms
with $E_z$ and $B_z$ in (5) and (6) become negligible, and GPM (4)
is reduced to (1). In other words, the additional terms in (4) can
be important in the case of classical radiation at the
intermediate and near zones only.

\section{Operator polarization matrix of quantum field}

Within the quantum domain, the field strengths in (5) and (6)
should be changed by the creation and annihilation operators of
photons with given polarization $E_{\mu},B_{\mu} \rightarrow
a_{\mu}$, $E^+_{\mu},B^+_{\mu} \rightarrow a^+_{\mu}$, where the
subscript $\mu$ takes the free different values corresponding to
the circular (transversal) polarizations with different helicities
and to the linear polarization in the radial (longitudinal)
direction \cite{5,7}. Leaving aside the known structure of the
position dependent mode functions, we can conclude that the
operator matrix $R({\vec r}) $ (3) corresponds to the
normal-ordered form in the creation and annihilation operators:
\begin{eqnarray}
R \rightarrow {\hat R}_{n}={\hat R}(a^+a). \nonumber
\end{eqnarray}
In addition, one can define the anti-normal operator matrix
\begin{eqnarray}
R \rightarrow  {\hat R}_{an}={\hat R}(aa^+)=FF^+ \nonumber
\end{eqnarray}
through the change of order of factors in the classical matrices
(5) and (6) and successive quantization. Although within the
classical picture $F^+F=FF^+$, the difference of the two operators
\begin{eqnarray}
{\hat R}_{an}-{\hat R}_n \equiv R_0 \label{7}
\end{eqnarray}
defines, by construction, the zero-point oscillations (ZPO) of the
electromagnetic field. In fact
\begin{eqnarray}
\langle 0|{\hat R}_{an}-{\hat R}_n|0 \rangle = \langle
0|R([a,a^+])|0 \rangle = R( \langle 0|[a,a^+]|0 \rangle ).
\nonumber
\end{eqnarray}
Due to the commutation relation $[a_{\alpha},a^+_{\beta}]=
\delta_{\alpha \beta}$, the difference (7) is represented by the
$c$-number matrix
\begin{eqnarray}
R^{(0)}= \left( \begin{array}{cc} [{\vec E},{\vec E}^+] & {\vec E}
\times {\vec B}^+-{\vec E}^+ \times {\vec B} \\ {\vec E}^+ \times
{\vec B} - {\vec E} \times {\vec B}^+ & P_{an}-P_{n} \end{array}
\right) \label{8}
\end{eqnarray}
The scalar term here describes the contribution of ZPO of the
electric field into the energy density since \cite{10}
\begin{eqnarray}
\langle 0|({\vec E}^+ +{\vec E})^2|0 \rangle =[{\vec E},{\vec
E}^+]. \nonumber
\end{eqnarray}
In turn, the off-diagonal vector terms in (8) describe ZPO  of the
Pointing vector, while the $(3 \times 3)$ matrix
\begin{eqnarray}
P^{(0)}=P_{an}-P_n \nonumber
\end{eqnarray}
gives ZPO of polarization \cite{11}.

It seems to be very interesting to compare ZPO of quantum
electromagnetic field in the two representations, i.e. for the
plane and spherical waves of photons. The energy of free field is
described by the following well-known formulas \cite{5}
\begin{eqnarray}
H_{plane}= \sum_{k, \sigma} \hbar \omega_k (a^+_{k \sigma}a_{k
\sigma}+1/2), \nonumber \\ H_{sph}= \sum_k \sum_{\lambda ,j,m}
(a^+_{k \lambda jm}a_{k \lambda jm}+1/2), \label{9}
\end{eqnarray}
for the plane and spherical waves of photons respectively. Here
$\sigma =1,2$ is the index of polarization of plane waves,
$\lambda$ labels the type of multipole radiation either electric
or magnetic, $j=1,2, \cdots ,$ is the angular momentum of photons,
and $m=-j, \cdots ,j$. The energy of ZPO in the whole volume of
quantization is provided by the averaging of (9) over the vacuum
state that gives
\begin{eqnarray}
H_{plane}^{(0)}= \sum_k \hbar \omega_k , \nonumber \\
H_{sph}^{(0)}= \sum_{\lambda} \sum_k \sum_{j=1}^{\infty} (2j+1)
\hbar \omega_k /2. \label{10}
\end{eqnarray}
At first sight, the two expressions in (10) are equivalent because
both give the infinite energy of the vacuum state. At the same
time, it is known that this infinity is inessential because any
measurement used to recognize ZPO implies an averaging over a
finite exposition time of detector \cite{12,13}. Such an averaging
is equivalent to a filtration, leading to a separation of a finite
transmission frequency band. It is clear that the second
expression in (10) strongly exceeds the first one in the case of
the same transmission frequency band. Even if we assume that
filtration separates only the electric dipole radiation ($\lambda
=E$, $j=1$), it is seen that
\begin{eqnarray}
H_{sph}^{(0)}/H_{plane}^{(0)}=3/2. \nonumber
\end{eqnarray}
This means that the measuring level of ZPO of multipole field
exceeds that of the plane waves of photons. This result follows
from the fact that the multipole radiation has much more quantum
degrees of freedom than the plane waves and each degree of freedom
contributes into the vacuum fluctuations. Since the plane wave
representation corresponds to the free field in empty space, while
the spherical wave representation takes into account the existence
of a singular point (emitter or absorber of the radiation field),
this means that the presence of a local source (atom) leads to an
increase of ZPO.

In a certain sense, the obtained picture of ZPO of spherical waves
of photons results from the spatial symmetry of solution of the
homogeneous wave equation used in the quantization \cite{5}. It
does not need any physical specification of the source considered
as a singular point for the spherical waves. This means that ZPO
described by $H_{sph}^{(0)}$ in (10) are independent on whether
the atom is used as a source or as a detector.

Let us note in this connection that possible influence of an atom
on the electromagnetic vacuum state in the absence of radiation
has been discussed in quantum electrodynamics for a long time
(e.g., see \cite{14}). In the usual picture, the so-called
$A^2$-term arising from the atom-field interaction is considered
is responsible for the renormalization of the vacuum fluctuations.
In the case under consideration, the effect is caused by the
geometry of space provided by the presence of atom independent of
the atom field interaction. Therefore, in a certain sense, the
effect is similar to the Casimir one \cite{15}.

Consider now the spatial structure ZPO.  Independent of the
representation, the density of ZPO can be described as follows
\begin{eqnarray}
H^{(0)}({\vec r})= \frac{k^2}{8 \pi} [{\vec A}({\vec r}),{\vec
A}^+({\vec r})], \label{11}
\end{eqnarray}
where ${\vec A}({\vec r})$ is the positive-frequency part of
vector potential at the point $\vec r$ and ${\vec E}({\vec
r})=-ik{\vec A}({\vec r})$ in the case of harmonic field. The mode
function of plane waves has the form $ \exp (i{\vec k} \cdot {\vec
r})$. Then, the commutator in the right-hand side of (11) is
independent of $\vec r$. In other words, ZPO of plane waves
manifest the spatial homogeneity.

In turn, the mode functions $V_{\lambda kjm \mu}({\vec r})$ of the
multipole radiation are represented by  certain combinations of
the spherical Bessel functions responsible for the radial
dependence and spherical harmonics describing the angular
distribution of radiation \cite{5,6}. Therefore, the right-hand
side of (11) is represented as follows
\begin{eqnarray}
H^{(0)}({\vec r})= \sum_{\lambda} \sum_{k,j,m, \mu} |V_{\lambda
kjm \mu}({\vec r})|^2. \label{12}
\end{eqnarray}
The explicit form of $V_{\cdots}({\vec r})$ can be found in
\cite{5,6,16}. It is a straightforward matter to show that (12) is
independent of the angular variables. Due to the properties of
spherical Bessel functions, the radial dependence of (12) shows an
inhomogeneous behaviour. Viz, ZPO of the multipole field are
concentrated in a certain vicinity of the source. At far
distances, they tend to the level predicted by the representation
of plane waves. The dipole ($j=1$) contribution into ZPO is shown
in Fig. 1. It is seen that inside the spherical region of the
radius $r_0 \sim 2/k= \Lambda / \pi$ ($\Lambda$ denotes the
wavelength), surrounding the atom, the level of multipole ZPO
strongly exceeds that of plane waves of photons.

Similar effect can be observed for ZPO of the polarization.
Consider as an illustrative example the first term in (4)
described by the matrix (5). It is seen that the corresponding
contribution into (8) is provided by the matrix with the elements
\begin{eqnarray}
P^{(0)}_{\mu \mu'}({\vec r})=k^2[A_{\mu}({\vec
r}),A^+_{\mu'}({\vec r})]. \nonumber
\end{eqnarray}
It is straightforward to calculate these elements in the same way
as (12). This yields the following. ZPO of polarization are
independent of the angular variables, while strongly depend on the
distance from the source. Then, ZPO of polarization are
concentrated in some vicinity of the source as well as (12).
Similar result can easily be obtained for (6).

\section{Conclusion}

Let us briefly summarize the obtained results. It is shown that
the general polarization matrix of an electromagnetic radiation
(4) can be constructed directly from the field-strength tensor
(2). This matrix is additive with respect to the contributions
coming from the electric and magnetic fields and quantitatively
describes the anisotrophy of the fields in terms of intensities of
radiation in different directions and corresponding phase
differences. In spite of the fact that (4) is examined here in the
case of plane and spherical waves of photons, the same matrix can
be used to specify the polarization of radiation in an arbitrary
geometry, for example in the case of a circular cylindrical
cavity. The quantization of electromagnetic field in such a cavity
was considered in \cite{17}.

By construction, the general polarization matrix is a local object
and the polarization of electromagnetic radiation may change from
point to point. Only in the case of plane waves, the polarization
is the spatially homogeneous property of the field (in the
direction of propagation).

Similar to the field energy, the polarization manifests ZPO which
can be calculated with the aid of the anti-normal and normal
ordered operator polarization matrices. ZPO of polarization define
the level of quantum noise in the polarization measurements. In
the case of plane waves of photons, the level of ZPO has the same
value at any point. The presence of a real local source (atom)
leads to an increase of ZPO above the level predicted by the model
of plane waves. This effect of concentration of ZPO near atom
corresponds to the near and intermediate zones. To stress the
importance of the obtained results, we note that the interatomic
distances in a number of modern experiments on engineered
entanglement in the system of trapped Ridberg atoms are of the
order of $r_0$ (e.g., see \cite{18} and references therein).
Therefore, there should be a strong vacuum noise which can
influence the precision of polarization measurements in such a
system. The above effect can also be considered in the context of
Casimir effect in atomic systems at short distances.

One of the authors (A.S.Sh.) would like to thank Professor J.H.
Eberly for useful discussions.

\begin{figure}
\caption{The radial dependence of the first term (at $j=1$) in
(12). The level of ZPO of the plane waves is shown by the dotted
line.}
\end{figure}

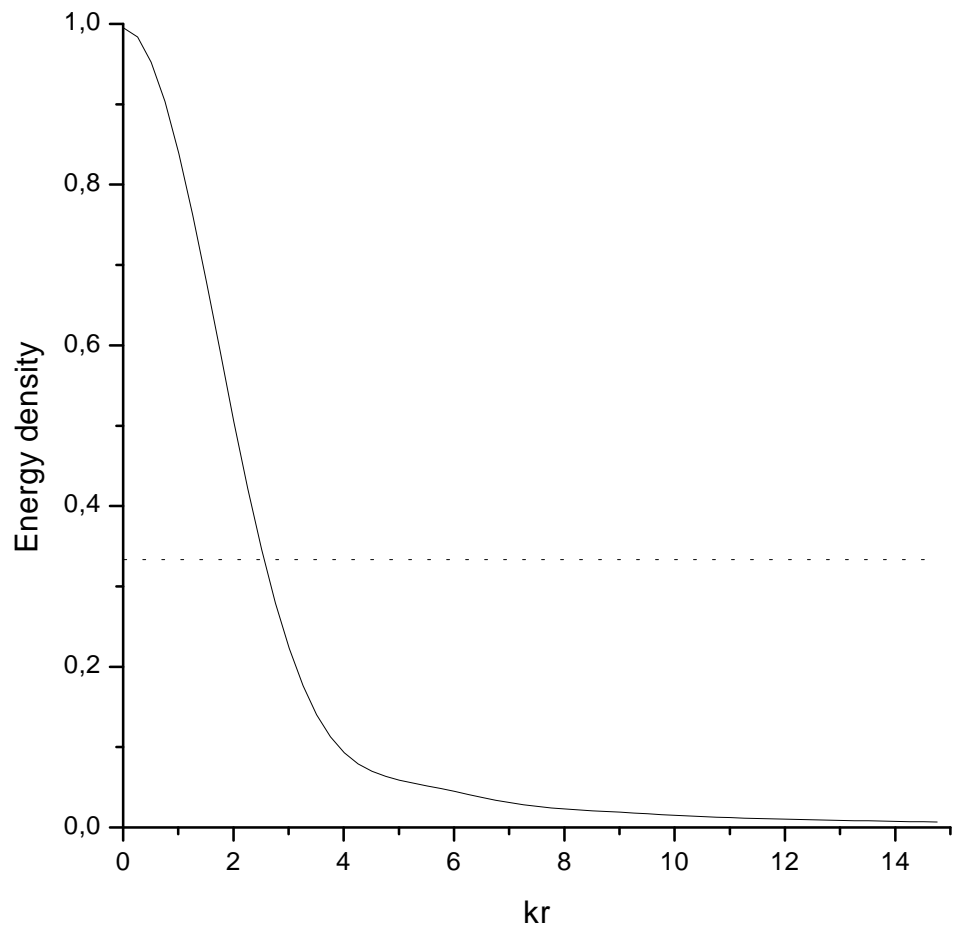
\end{document}